\documentclass[showpacs,preprintnumbers,amsmath,amssymb,aps,onecolumn, preprint]{revtex4}

\usepackage{graphicx}% Include figure files
\usepackage{dcolumn}
\usepackage{bm}
\usepackage{epsfig}
\usepackage[ansinew]{inputenc}
\begin{document}

\title{Phase transitions in number theory: from the Birthday Problem to Sidon Sets}
% Force line breaks with \\

\author{Bartolo Luque$^{1}$, Iv\'{a}n G. Torre$^1$, and Lucas Lacasa$^{2}$}
\email{lucas.lacasa2@gmail.com}
\affiliation{$^1$Departamento de Matem\'{a}tica Aplicada y Estad\'{i}stica, ETSI Aeron\'{a}uticos,  Universidad Polit\'{e}cnica de Madrid, Spain\\
$^2$School of Mathematical Sciences, Queen Mary, University of London, Mile End Road, London E1 4NS, UK}%

\date{\today}
% It is always \today, today, % but any date may be
% explicitly specified


\begin{abstract}
In this work, we show how number theoretical problems can be fruitfully approached with the tools of statistical physics. We focus on g-Sidon sets, which describe sequences of integers whose pairwise sums are different, and propose a random decision problem which addresses the probability of a random set of $k$ integers to be g-Sidon. First, we provide numerical evidence showing that there is a crossover between satisfiable and unsatisfiable phases which converts to an abrupt phase transition in a properly defined thermodynamic limit. Initially assuming independence, we then develop a mean field theory for the g-Sidon decision problem. We further improve the mean field theory, which is only qualitatively correct, by incorporating deviations from independence, yielding results in good quantitative agreement with the numerics for both finite systems and in the thermodynamic limit.
Connections between the generalized birthday problem in probability theory, the number theory of Sidon sets and the properties of q-Potts models in condensed matter physics are briefly discussed.
\end{abstract}

\maketitle %%%
\section{Introduction}
From the celebrated work on the critical behavior of random boolean satisfiability (random k-SAT) \cite{FIRST, selman}, we have seen how statistical physics and computer science have built bridges between each
other, as it has been recognized that the theory underlying optimization problems \cite{bookSAT} and the physics of disordered systems \cite{spin_glass} shares many similarities \cite{mertens2, moore}. In computer science, one can investigate the situations under which decision problems are satisfiable, and study the geometry and accessibility to those solutions. Decision problems can indeed by interpreted under a statistical physics formalism, where the cost function to be minimized (in random k-SAT, this is for instance the number of violated constraints) relates to the Hamiltonian of a disordered system at zero temperature: in this latter situation, the physical system tries to adopt the ground state or minimal energy configuration, only reachable in some circumstances, while in others frustration (unsatisfiability) can develop. In the last years, an exciting multidisciplinary environment has witnessed the efforts of describing optimization problems within the physics of disordered problems, including replica-symmetry-breaking solutions in combinatorial problems \cite{spin_glass} or the description of phase transitions (threshold phenomena) in decision problems \cite{bookSAT,monasson,monasson2,mertens}. In a nutshell, the mutual interchange of approaches and techniques from physics to computer science and viceversa have proved to be a valuable input in both fields \cite{moore}.\\

\noindent Here we explore the possibility of further extending this fruitful relation to number theory. Can arithmetics and number theory be seen as a natural system, subject to scientific scrutiny much in the form physics observes physical reality? Can numbers be considered as units that interact locally according to some arithmetic properties and hence be amenable to a statistical mechanics description? Several works following this epistemological approach range from the reinterpretation of the nontrivial Riemann zeta zeroes as the eigenspectrum of a quantum Hamiltonian \cite{berry} to the onset of phase transitions in number partitioning problems \cite{mertens, mertens2}. Within complexity science, a fundamental question is to find the minimal amount of ingredients a system needs to possess to evidence emergent behavior. Number theoretic systems can indeed be thought of as the utterly purest, unadorned and simplest models where complexity may develop \cite{PRL}, and therefore constitute a privileged playground for scientific research. In this letter we show how number theoretical problems are susceptible to be approached from a computer science and statistical physics perspective, and indeed show nontrivial emergent properties. For concreteness, we focus on the arithmetics of Sidon sets \cite{abbot} and show that this number theoretical statement is susceptible to be treated as a random Constraint Satisfaction Problem (rCSP), and show links with both the statistical physics of disordered systems and with some classical problems in probability theory. In doing so, we will unveil the onset of a phase transition in the satisfiability of such system -a so called zero-one law in the mathematical jargon \cite{zero1,zero2}-, which we will show is analytically tractable.\\

\section{random g-Sidon: a number-theoretic decision problem}

In number theory, a set of $k$ different positive integers ${\cal X}=\{S_1,S_2,...,S_k\}$, is a so called Sidon set \cite{abbot} if all the sums of two elements $S_i+S_j$ from the set (where $i,j=1,...,k$) are different (except when they coincide because of commutativity). For example, $\{1, 2, 5, 10, 16, 23\}$ is a Sidon set, whereas $\{1, 3, 7, 10, 17, 23\}$ is not Sidon since $1 + 23 = 7 + 17$. Sidon sets recurrently appear in different areas of mathematics including Fourier analysis, group theory or number theory. An extension of Sidon sets allows in the definition for $g$ repetitions, accordingly, ${\cal X}$ is called $g-$Sidon provided that any sum of two elements $S_i+S_j$  is \emph{repeated} at most $g-1$ times (note that when $g=1$, a $g$-Sidon set reduces to a Sidon set). From its first beginnings \cite{erdos}, number theorists were interested in the extremal properties of Sidon sets, concretely in calculating upper and lower bounds of the maximal size of $g$-Sidon sets formed from an integer interval $[1,M]$ for diverging values of $M$, a topic and focus which still has an intense research activity \cite{ruzsa,cilleruelo,yu}. Less attention (if any) has been paid in the onset of zero-one laws in such systems \cite{zero3}. Here we show that a statistical physics approach is helpful in this task. We begin by recasting the concept of Sidon sets and propose a random constraint satisfaction problem (rCSP) approach as it follows: let $[1,M]$ be a pool of positive integers and extract at random from this pool a set of $k$ different numbers ${\cal X}=\{S_1,S_2,...,S_k\}\subset [1,M]$. Which is the (satisfaction) probability $P_g(k,M)$ that ${\cal X}$ is a $g$-Sidon set? We will call this rCSP the random $g$-Sidon decision problem. First, notice that this problem can be rephrased in a statistical physics formalism in the following terms: given the initial set ${\cal X}$, proceed to build a secondary set $\tilde{{\cal X}}=\{S_i+S_j\}_{i,j=1,...,k}=\{\tilde{S}_n\}_{n=1,...,k(k+1)/2}\subset [2,2M]$ formed by all the sums of two elements from the set ${\cal X}$. As the sums $S_i+S_i$, $i=1,...,k$ are allowed,  $\tilde{{\cal X}}$ will be formed by $k(k+1)/2$ positive integers between $2$ and $2M$. $\tilde{{\cal X}}$ is therefore the output of a Sidon set ${\cal X}$ if every number in $\tilde{S}_n$ distinct. Let us define the number of matches of $\tilde{S}_n$ in $\tilde{{\cal X}}$ as:
\begin{equation}
h_n= \sum_{m=1,m\neq n}^{\frac{k(k+1)}{2}}\delta[\tilde{S}_n-\tilde{S}_m],
\label{hi}
\end{equation}
where the Kronecker function $\delta[x]=1$ if $x=0$ and $0$ otherwise.
Then, for $g=1$, we can define the following function ${\cal H}$:
\begin{equation}
{\cal H}(g=1,k,M,\tilde{{\cal X}})= \sum_{n=1}^{\frac{k(k+1)}{2}}h_n.
\label{Hamiltonianog1}
\end{equation}
${\cal H}$ computes 'the degree of Sidonlikeness' of a configuration $\{S_i\}$, i.e. the total number of matches in $\{\tilde{S}_n\}$. Indeed, ${\cal H}=0$ for Sidon sets while ${\cal H}>0$ for non-Sidon sets. Stated as a rCSP, ${\cal H}=0$ and ${\cal H}>0$ distinguish the satisfiable and unsatisfiable phases respectively, such that satisfiability is reached for configurations that minimizes this  function.
%Note also that although ${\cal H}$ scales quadratically with respect to $k$, in principle this function scales linearly with the coarse-grained magnitude $\|\tilde{\textbf{S}}\|$, being therefore an extensive quantity in the number of variables.
 ${\cal H}$ can be seen as the physical internal energy of the statistical mechanics system $\tilde{{\cal X}}$, where each configuration of the variables $\{\tilde{S}_n\}$ is a given microstate with energy ${\cal H}$. Each of the $k(k+1)/2$ 'spins' can take discrete values in $[2,2M]$.
 If random fluctuations of the spin values were allowed, the equilibrium properties of this system at temperature $T$ would be given by the Boltzmann measure in $[2,2M]$ $\mu(\tilde{\textbf{S}})=\exp[-\beta{\cal H}(\tilde{\textbf{S}})]/Z$, where $\beta\sim 1/T$, and $Z$ is the normalization (partition) function. In general, the system will be in the phase (Sidon/non Sidon) that minimizes the Helmholtz free energy $F={\cal H}-TS$.
 Since no random fluctuations of the values of the variables are allowed in our system, the system is to be considered at zero temperature, and variables will try to occupy the ground state energy, that is, the configuration that minimizes ${\cal H}$, being this the satisfiable phase if $\min[ {\cal H}]=0$, and getting frustrated in the unsatisfiable one if the minimum energy state available is larger than zero.\\
The case of $g$-Sidon sets ($g>1$) is slightly more involved:
\begin{equation}
{\cal H}(g,k,M,\tilde{{\cal X}})=\sum_{n=1}^{\frac{k(k+1)}{2}}(h_n-g+1)\theta[h_n-g],
\label{Hamiltonianogen}
\end{equation}
where the Heaviside step function $\theta[x]=0$ if $x<0$ and $\theta[x]=1$ if $x\geq0$. Hence, $g$-Sidon sets and non $g$-Sidon comply ${\cal H}=0$ and ${\cal H}> 0$ respectively.\\

\begin{figure}[t]
\leavevmode \epsfxsize=12 cm \epsffile{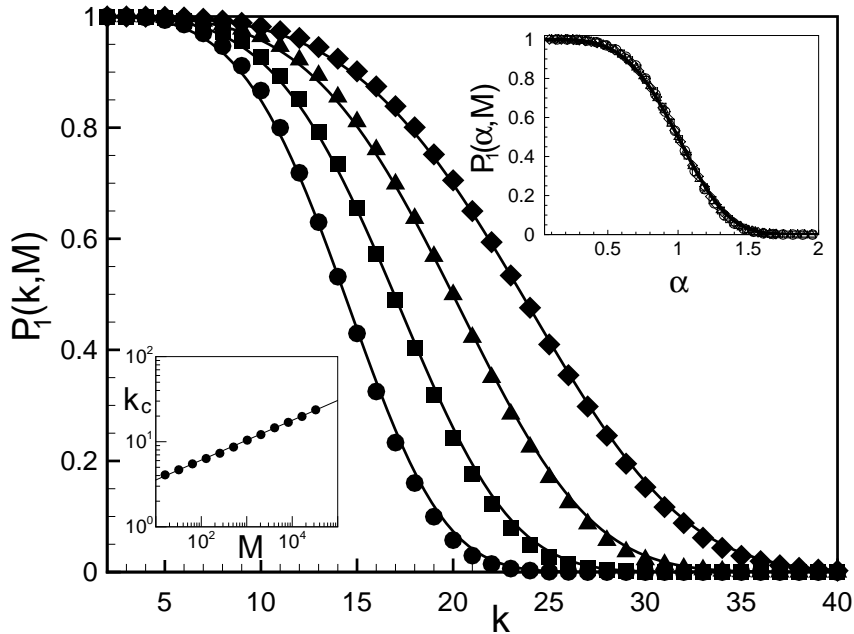}
\caption{(Central Panel) Numerical simulations of satisfaction probability $P_1(k,M)$ as a function of $k$, for $M=2^{12}$ (circles), $M=2^{13}$ (squares), $M=2^{14}$ (triangles), $M=2^{15}$ (diamonds), averaged over $10^4$ realizations. Solid lines are the prediction of equation \ref{3}. (Inset bottom panel) Log-log plot of the transition point $k_c(M)$, defined as $P_1(k_c,M)=1/2$, showing the scaling $k_c \sim M^{1/4}$. (Inset top panel) Collapsed satisfaction probability $P_1(\alpha,M)$, for $\alpha = k/k_c$, finding a continuous sigmoidal function independent of $M$. Solid line is a prediction of the theory $P_1(\alpha)=2^{-\alpha^4}$ (see the text).}
\label{g1}
\end{figure}

\section{Numerical results: from Crossover to phase transitions}
\noindent The order parameter which naturally associates with sat/unsat phases in the random $g$-Sidon problem is the satisfaction probability $P_g(k,M)$, that describes the probability of a randomly extracted set of $k$ elements from $[1,M]$ to be $g$-Sidon, i.e., to fulfill $\min [{\cal H}]=0$. We will consider $k$ as the control parameter and in what follows we firstly explore numerically the behavior of $P_g(k,M)$ as a function of $M$ and $g$. The behavior for $g=1$, as a result of (ensemble-averaged) Monte Carlo simulations, is shown in figure \ref{g1}. First, notice that the transition between satisfiability ($P_g(k,M)\approx 1$) and unsatisfiability ($P_g(k,M)\approx 0$) occurs at increasing values $k(M,g)$. In order the control parameter to be intensive, we rescale it as $\alpha=k/k_c(M,g)$, where we make use the standard in percolation theory and set $P_g(k_c,M)=1/2$. In the bottom inset panel of the same figure we show the explicit dependence $k_c(M)$, which we find agrees with the expression $k_c(M)\sim M^{1/4}$. Notice that $k$ is not actually extensive (linear) in $M$ and therefore the ratio $k/M$ typically used in the satisfiability theory \cite{bookSAT} does not work here.\\

\noindent In the upper inset panel of the same figure we also show the behavior $P_1(\alpha,M)$ as a function of $\alpha$, for different pool cardinals $M$. The collapse of the order parameter under a single smooth curve points out that the behavior is independent of $M$, which means that the pool's size does not play any relevant role, and $P_1(k,M)_{\overrightarrow{k\mapsto\alpha}} P_1(\alpha)$. Also, such transition is smooth both for finite sizes and in the thermodynamic limit ($k\rightarrow \infty$, $M\rightarrow \infty$, $\alpha$ finite), that is, the system seems to evidence a simple crossover and no threshold phenomenon occurs.\\

\begin{figure}[t]
\leavevmode \epsfxsize=12 cm \epsffile{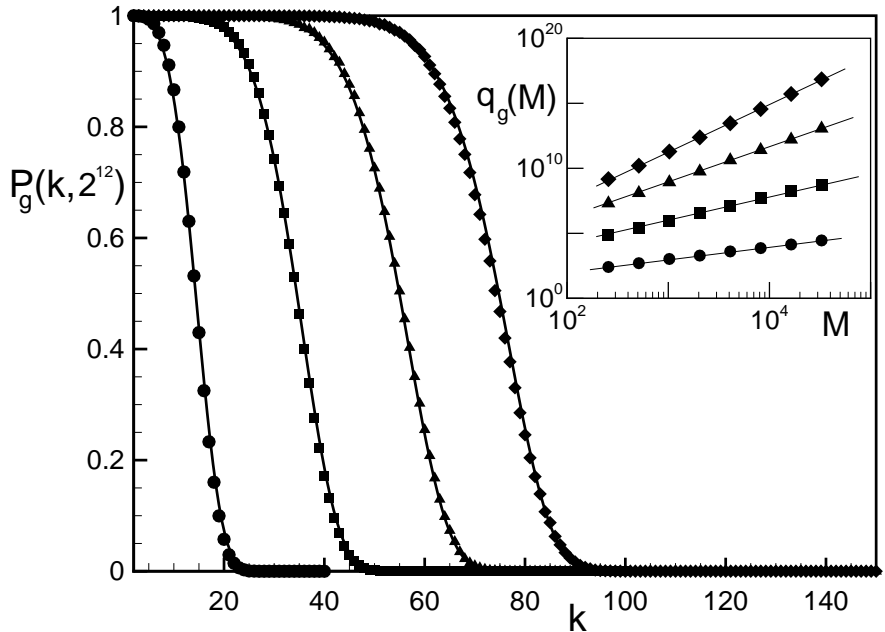}
\caption{(Central Panel) Numerical simulations of satisfaction probability $P_g(k,M=2^{12})$ as a function of $k$, for $g=1$ (circles), $g=2$ (squares), $g=3$ (triangles), and $g=4$ (diamonds) (Monte Carlo simulations averaged over $5\cdot10^4$ realizations). Solid lines are the prediction of the theory (see the text). (Inset panel) Numerical scalings of $q_g(M)$ for different values of $g$, finding in each case a power law relation albeit with different slopes, suggesting the two-variable scaling $q_g(M)=A(g)M^{r(g)}$. The expression $q_g$ is related to $k_c$ through the transformation $q_g(M)\equiv k_c^{2(g+1)}/((g+1)!2^{2g+1}\log 2)$ (see the text). The specific shapes of $A(g)$ and $r(g)$ are plotted the insets of Fig. \ref{fig3}.}
\label{fig2}
\end{figure}

\begin{figure}[t]
\leavevmode \epsfxsize=12 cm \epsffile{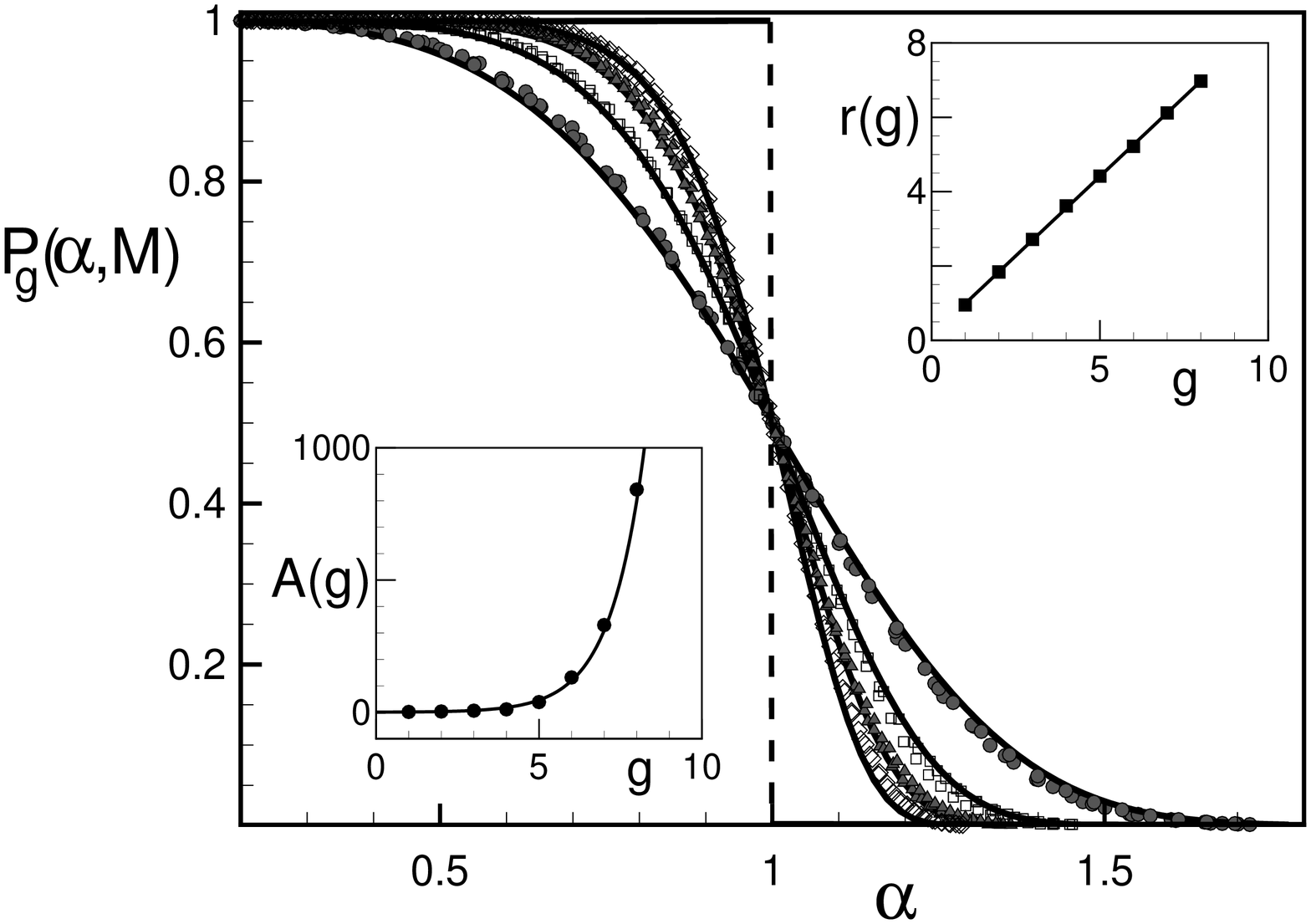}
\caption{Rescaled satisfaction probability $P_g(\alpha,M)$ as a function of $\alpha$, for several values of $g=1$ (grey circles), $g=2$ (squares), $g=3$ (grey triangles), and $g=4$ (diamonds), and $M=2^{12}, 2^{13}, 2^{14}, 2^{15}$. For each $g$, there is a universal collapse curve $P_g(\alpha)$ which turns to be independent of $M$, however, such universal curves get sharper around $\alpha=1$ for increasing values of $g$, suggesting the onset of a phase transition in the limit $g\rightarrow\infty, \alpha$ finite. Solid lines for finite $g$ and the Heaviside function for $g\rightarrow\infty$ are predictions of the theory (see the text). (Inset top panel) Linear fit of the correcting exponent $r(g)\approx0.85g+0.13$ ($R^2=0.9998$).(Bottom inset panel) Exponential fit of the correcting factor $A(g)\approx 0.4\exp(0.95g)$ ($R^2=0.9985$).}
\label{fig3}
\end{figure}

\noindent In order to cast light in the effect of $g$, in Fig.\ref{fig2} we show the numerical results of $P_g(k,M)$ for a concrete pool size $M=2^{12}$, for different values of $g$. Again, the transition point $k_c$ shows a dependency not only with $M$ but also with $g$ (see the inset panel). When the control parameter is properly made intensive, and at odds with the phenomenology for $g=1$, we find that the transition \emph{sharpens} for increasing values of $g$. This result is further confirmed in Fig. \ref{fig3}, where we plot the behavior of $P_g(\alpha,M)$ for different values of $M$ and $g$. For finite $g$ the satisfaction probability adopts again a universal $M$-independent sigmoidal curve $P_g(k,M)_{\overrightarrow{k\mapsto\alpha}} P_g(\alpha)$, although this curve gets sharper around $\alpha=1$ as $g$ increases. This sharpening further suggests that $g$ plays the role of an \emph{effective} system size, such that the crossover that takes place for finite $g$ seems to develop into a true phase transition in the limit of $g\rightarrow \infty$. This is a genuinely counterintuitive result having in mind that in decision problems the apparent system's size is usually related to pool's size, here $M$, which in our case is an irrelevant variable.

\section{Analytical developments}

In what follows we support this phenomenology with some analytical calculations. Incidentally, note at this point that if $\tilde{S}_i$ were drawn uniformly from $[2,2M]$, the 1-Sidon problem would be equivalent to the celebrated Birthday Problem \cite{feller}, a standard in probability theory that calculates the probability that, in a set of $k$ randomly chosen people, not a single pair will share the same birthday (with a year containing $N=2M-1$ days). In that case equation \ref{Hamiltonianog1} would also be equivalent to the Hamiltonian of a $N$-Potts model in a mean field approximation (no explicit space) widely used in solid state physics \cite{potts}. Similarly, if again $\tilde{S}_i$ were drawn uniformly, the $g$-Sidon problem would reduce in the secondary description to the so-called generalized Birthday problem \cite{Generalized birthday}, which calculates the probability that if $k$ people are selected at random, $g+1$ people won't have the same birthday (or alternatively, the probability that at most $g$ people will share the same birthday). In our problem $\tilde{S}_n$ are the birthdays and $2M-1$ the days in one year. Since $\tilde{S}_n$ are non-uniformly sampled in $[2,2M]$ (as they are the result of the sum pairs in $\{S_i\}$), we will only take the Birthday problem/Potts model as naive approximations of the random $g$-Sidon problem.\\

%A straightforward combinatorial argument lead us to express the satisfaction probability $P_1(k,M)$ as
%\begin{equation}
%P_1(k,M)=\frac{\prod_{i=0}^{k(k+1)/2-1} 2M-i}{(2M)^{k(k+1)/2}}\approx \exp(-k^4/16M).
%\label{1}
%\end{equation}
%The behavior in the thermodynamic limit can be straightforwardly extrapolated from eq. \ref{1} if we only assume the numerical evidence $k_c(M)\sim M^{1/4}$, finding $P_1(\alpha,M)=2^{-\alpha^4}$: a continuous sigmoidal curve which represents a simple crossover phenomenon (see the top inset panel of figure \ref{g1} for a comparison between theory and numerics).\\
%The second step is to generalize eq. \ref{1} $\forall g$, which is more involved. Working again within the secondary description,
\noindent Suppose that a year has $N$ days quote $P_g(n,N)$ the probability that no $g+1$ people, of $n$ people selected at random, have the same birthday. Then the following recursive relation holds approximately:
$$P_g(n+1,N|n,N)^N\approx P_{g-1}(n,N).$$
Using Bayes' theorem and taking logarithms in the preceding equation, we find
$$\log P_g(n+1,N)-\log P_g(n,N)=\frac{1}{N}\log P_{g-1}(n,N)$$
Now, for sufficiently large $N>>n$,  the lhs in the latter expression is a first order discretization of $\partial \log P_g/\partial n$. In the continuum limit this yields a partial differential equation for the evolution of $P_g$
$$\frac{\partial P_g(n,N)}{\partial n}=\frac{P_g(n,N)}{N}\log P_{g-1}(n,N)$$
that along with initial condition $P_g(0,N)=1$ for all $g$ has the following solution
\begin{equation}
P_g(n,N)=\exp\bigg(-\frac{n^{g+1}}{(g+1)!N^g}\bigg).
\label{2}
\end{equation}
%Interestingly, note that for $g>1$, the calculation of $P_g(k,M)$ in the naive case of uniform sampling can be reduced to the so-called generalized birthday problem \cite{Generalized birthday}.
The generalized Birthday problem has an analytically unmanageable closed-form expression \cite{Generalized birthday}, that nonetheless has been treated asymptotically by some authors \cite{mendelson, henze}. We find noticeable that the asymptotic solutions to the problem, derived using slightly sophisticated combinatorial and statistics techniques, agrees with our expression in eq.\ref{2}, obtained following a straightforward argument.\\

\noindent If we assume in our problem that the values $\tilde{S}_n$ are uncorrelated and result of independent trials in $[2,2M]$, $P_g(k,M)$ results from the change of variables:  $N\rightarrow 2M$, $n\rightarrow k(k+1)/2\approx k^2/2$:
\begin{equation}
P_g(k,M)=\exp\bigg(-\frac{k^{2(g+1)}}{(g+1)!2^{2g+1}A(g) M^{r(g)}}\bigg),
\label{3}
\end{equation}
where we have formally substituted $M^g$ with $A(g)M^{r(g)}$ because the $n=k(k+1)/2$ variables are correlated in our system. In order to quantitatively compare our theory with finite-size numerics, we shall express higher order deviations from this equation introducing a correcting exponent $r(g)\neq g$, whose first order perturbative expansion reads $r(g)=r_0+r_1g$, and similarly for the normalizing factor $A(g)$. The concrete values of the free parameters are then found using a simple self-consistent argument, imposing that the correct scalings $k_c(M,g)$ shall be found at $P_g(k_c,M)=1/2$. After a little algebra we find that $q_g(M)\equiv k_c^{2(g+1)}/((g+1)!2^{2g+1} \log 2)= A(g)M^{r(g)}$, where the specific fits for $A(g)$ and $r(g)$ are shown in the inset panels of Figs. \ref{fig3}. This expression reduces to $k_c\sim M^{1/4}$ for $g=1$, on agreement with previous numerical evidence (inset panel of Fig. \ref{g1}). The predicted values of $P_g(k,M)$ are accordingly plotted in solid lines along with the numerics in Figs. \ref{g1}, \ref{fig2}, \ref{fig3} for different values of $M$ and $g$, showing an excellent agreement in every case.\\ %Note also that the accurate expression for $P_{g=1}(k,M)$ is qualitatively equivalent to equation \ref{1} found using very basic combinatorics, although there is a slight quantitative difference in the normalizing factor that accounts for the nonuniform sampling bias.\\

\begin{figure}[t]
\leavevmode \epsfxsize=10 cm \epsffile{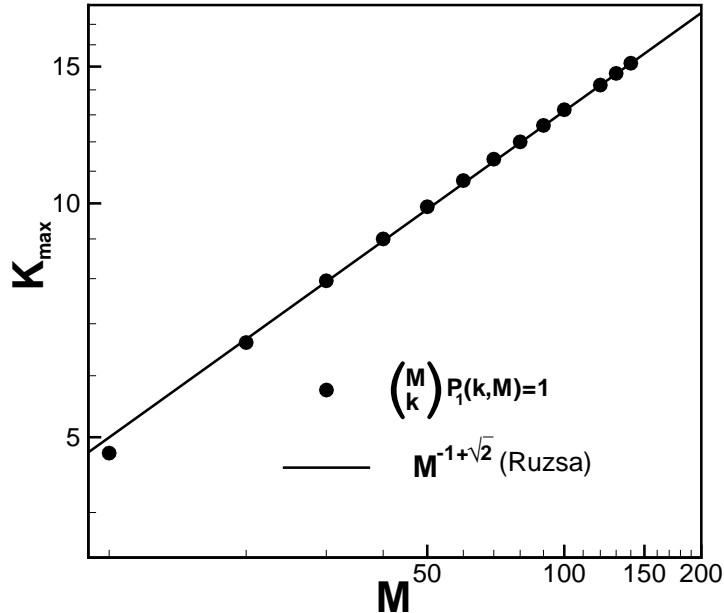}
\caption{(Dots) Maximal size of a 1-Sidon set as a function of $M$, as predicted by the phenomenological theory underlying the random Sidon decision problem. (Solid line) Maximal 1-Sidon set bound given by Ruzsa \cite{ruzsa}.}
\label{Rusza}
\end{figure}

\noindent Incidentally, our theory also predicts that the maximal size $k_{\textrm{max}}$ of a $g$-Sidon set -a classical question in number theory- should satisfy ${M \choose k_{\text{max}}}P_g(k_{\text{max}},M)=1$, whose leading order for $g=1$ is $k_{\text{max}}={\cal O}(M^{\sqrt 2 -1})$, on agreement with a recent theorem by Ruzsa \cite{ruzsa} (see Fig.\ref{Rusza}).\\

\noindent Finally, as a function of the intensive control parameter $\alpha=k/k_c$, eq.\ref{3} reduces to
\begin{equation}
P_g(\alpha,M)= 2^{-\alpha^{2(g+1)}}.
\end{equation}
It is important to highlight that this law holds independently of the concrete values of $A(g)$ and $r(g)$ -that is, it is a direct consequence of the theory-, and also holds without needs to impose taking the limit $M\rightarrow \infty$ (see Fig \ref{fig3}). The solution that we obtain is a universal sigmoid function for finite $g$ (in the case of $g=1$, the curve is $P_1(\alpha)=2^{-\alpha^{4}}$, on excellent agreement with the numerics in the inset panel of Fig. \ref{g1}). This sigmoid sharpens for increasing values of $g$, and in the thermodynamic limit $g\rightarrow \infty$ (what necessarily implies $k\rightarrow \infty$), $\alpha$ finite a zero-one law emerges
\begin{equation}
P_{\infty}(\alpha,M) =
\left\{
\begin{array}{rcl}
     1 & \textrm{if} & \alpha < 1
  %\\ 1/2 & \textrm{if} & \alpha = 1
  \\ 0 & \textrm{if} & \alpha >1,
\end{array}
\right.
\label{heaviside}
\end{equation}
 That is, whereas the decision problem only evidences a crossover for all finite $g$, this transition indeed becomes abrupt and converts to a true phase transition in the thermodynamic limit.

\section{Conclusion}
To conclude, we have shown how the methods and focus of statistical physics and theoretical computer science can be fruitfully applied in the realm of number theory. We have found and described, both numerically and analytically, a previously unnoticed phase transition within the properties of $g$-Sidon sets, with the exotic peculiarity that $M$ -the analog of the number of possible values of each spin, for instance $q$ in the $q$-Potts model- does not play any relevant role in the onset of the phase transition, while the finite-size role is played here by the combinatorial parameter $g$. The extension of these approaches to other number theoretical problems, and the establishment of new links between these fields are important open problems to be further addressed.

\end{document}